# Comments on 'Combining Switched TMAs and FDAs to Synthesize Dot-Shaped Beampatterns'

Mahdi Fartookzadeh

The idea of frequency diverse array (FDA) has been invigorated in the recent years by supposing to have the ability of stopping the electromagnetic wave propagation and producing secretive direct connections between distant points. However, it was based on a theoretical mistake in a series of published papers beginning from 2014. In fact, the issue was not very clear since 2018; hence the researchers should be very careful in relying on the papers of this topic. Unfortunately, although the authors of the title paper [1] admitted the time variance of the beampatterns produced by FDAs, they suggested that it can be overcome by imposing a restriction based on previously published papers. This is while it has been proven that a time-invariant focusing beampattern cannot be produced by using any method in the farfield [2, 3]. Furthermore, it has been proven that the time variance of the beampattern exactly follows its range variance, meaning that it is not possible to construct an array with time-invariant range-variant beampattern, in any way including the frequency diverse array (FDA) technique [4].

To be more specified about the misconception in [1], it should be noted that all the calculations in [1] are only valid for the farfield region by assuming $r_n \approx r - nd\sin\theta$ in Eq. (7) of [1]. In addition, it is noteworthy that the condition in Eq. (13) of [1] only makes the pulse duration so small that the angle change of the focusing point is negligible on it. Thus, it is not related to the movement of the focusing point through range direction. Hence, the beampatterns in Fig. 4 of [1] are only at an instant of time and they move through the range direction with the speed of light. For example, if we assume that the beampattern in Fig. 4 (a) of [1] is plotted at $t = t_0$, the focusing point will move through the range dimension by $c\Delta t$ at the time $t = t_0 + \Delta t$, where $c$ is the speed of light.

On the other hand, if we admit the time-variance of the beampatterns, it has been indicated that the concept of time-variant focusing point is exactly the same with the pulse transmission by using traditional phased arrays [5]. For example, the beam collection efficiency (BCE) of the 19-element FDA proposed in [1] is compared with BCEs of a regular 19-element phased array with Gaussian and rectangular pulses in Fig. 1. It can be observed that the same BCE (without any sidelobes) can be obtained by using a Gaussian pulse with full duration at half maximum (FDHM) of 16.7 μs, yielding a Gaussian pattern on the range dimension with the full width at half maximum (FWHM) of 5 km. The pattern width can be controlled by changing the FDHM of the input signal without needing to increase the number of array elements. Also, the beampattern shape can be controlled easily by controlling the input signal such as the example of a rectangular pulse in Fig. 1. It should be reaffirmed that all the beampatterns in Fig. 1 are instantaneous and they are moving with the speed of light. Finally, while the time at which the beampattern has been produced by the phased array is very obvious [5], a question arises about the time at which the beampattern has been produced by the FDA.

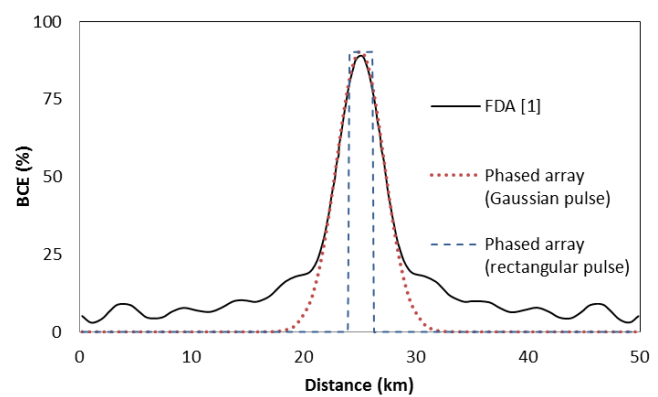

Fig. 1. BCE of the 19-element FDA proposed in [1] compared with BCEs of a regular 19-element phased array with Gaussian and rectangular pulses.

M. Fartookzadeh is with Department of Electrical and Computer Engineering, Malek Ashtar University, P. O. Box 1774-15875, Tehran, Iran (Mahdi.fartookzadeh@gmail.com).